\documentclass[runningheads]{llncs}
\usepackage{graphicx}
\usepackage{amsmath}
\usepackage{amssymb}
\usepackage{subfigure}
\usepackage[misc]{ifsym}
\begin{document}
\title{6VecLM: Language Modeling in Vector Space for IPv6 Target Generation}
%
%
\author{}
\author{Tianyu Cui\inst{1,2} \and Gang Xiong(\Letter) \inst{1,2}\and
Gaopeng Gou\inst{1,2}\and Junzheng Shi\inst{1,2}\\ \and Wei Xia\inst{1,2}
}

\authorrunning{T. Cui et al.}
%
\institute{}
\institute{Institute of Information Engineering, Chinese Academy of Sciences \and
School of Cyber Security, University of Chinese Academy of Sciences\\
Beijing, China\\
\email{ \{cuitianyu, xionggang, gougaopeng, shijunzheng, xiawei\}@iie.ac.cn}}
\maketitle              
\begin{abstract}
Fast IPv6 scanning is challenging in the field of network measurement as it requires exploring the whole IPv6 address space but limited by current computational power. Researchers propose to obtain possible active target candidate sets to probe by algorithmically analyzing the active seed sets. However, IPv6 addresses lack semantic information and contain numerous addressing schemes, leading to the difficulty of designing effective algorithms. In this paper, we introduce our approach 6VecLM to explore achieving such target generation algorithms. The architecture can map addresses into a vector space to interpret semantic relationships and uses a Transformer network to build IPv6 language models for predicting address sequence. Experiments indicate that our approach can perform semantic classification on address space. By adding a new generation approach, our model possesses a controllable word innovation capability compared to conventional language models. The work outperformed the state-of-the-art target generation algorithms on two active address datasets by reaching more quality candidate sets.

\end{abstract}
\section{Introduction}
Host discovery has always been a vital research method in the field of network measurement. By exploiting the ability of modern hardware and connectivity, tools like Zmap \cite{durumeric2013zmap} and Masscan \cite{graham2014masscan} have been able to complete the exploration of the global IPv4 address space, which has fundamentally enhanced the ability of researchers to conduct wide-ranging assessments of Internet services.

However, as has long been recognized, IPv6’s much larger address space \cite{deering2016internet} renders exhaustive probing completely infeasible. A recently proposed solution is to design a target generation algorithm \cite{ullrich2015reconnaissance,foremski2016entropy,murdock2017target} to generate a candidate set that may be active. Systems are required to analyze the potential distribution characteristics of the active address set and infer the target clustering area. The design of effective analysis algorithms directly determines the ability of model learning and the quality of generated candidate sets.

While prior work has obtained a preliminary understanding of active addresses distribution \cite{plonka2015temporal,gasser2018clusters}, the results commonly lack interpretability because IPv6 address consisting entirely of digits misses semantics \cite{carpenter2014significance}, conditioning our inability to infer active addresses using sequence relationships. The reason mainly comes from numerous customizable IPv6 addressing schemes \cite{thomson1998ipv6,narten2001privacy}. The complexity of address composition causes difficulty in algorithmic inferences.

The representative target generation algorithms include Entropy/IP \cite{foremski2016entropy} and 6Gen \cite{murdock2017target}, which are designed based on human observation and assumptions on network data. Human intervention may result in the algorithm overly dependent on experience and lose adaptability to the data set. The question of how to push the candidate set generated by the algorithm from quantity to quality remains.

To address these problems, we consider a new approach employing deep learning to facilitate effective IPv6 target generation. Word embedding \cite{mikolov2013efficient} and language modeling \cite{bengio2003neural,mikolov2010recurrent,grave2016improving,dauphin2017language} are a critical component of systems that require modeling long-term dependency, with successful applications such as summarization and machine translation. By word-to-vector space mapping, word vectors expose the semantic relationships between various words. Language models can estimate the probability distribution of a sequence of words by supervised learning. Based on these principles, we propose to construct an IPv6 vector space with a certain degree of semantic relationship. Through learning the semantic, a language model can autonomously infer the components of active addresses to generate more effective candidate results.

Conventional language models are used to model deterministic sequence dependency. The predicted sequence results are basically consistent with the original data set. In the target generation work, language models are required an innovation to satisfy creative sequence generation.

In this paper, we develop a new concrete instantiation of the target generation algorithm 6VecLM through deep learning, which includes two mechanisms IPv62Vec and Transformer-IPv6. IPv62Vec maps the entire active address space to a semantic vector space, where addresses with similar sequences will be classified into the same cluster. Semantic address vectors will be learned by Transformer-IPv6 to implement IPv6 language modeling. By modeling with a Transformer network  \cite{vaswani2017attention}, our work can comprehensively consider multiple sequence relationships and generate creative and semantically similar sequences to the data set. To serve the generative task, we decide to employ a new generation approach based on cosine similarity and softmax temperature \cite{muller2019does} to substitute the probability prediction in language models. Through choosing various sampling strategies, the model can generate expected and creative host targets.

\textbf{Contributions:} Our contributions can be summarized as follows:

\begin{itemize}
\item[1)] We explored the construction of IPv6's semantic space for the first time. IPv62Vec can effectively cluster the active address space into several classes.
\item[2)] We designed a new target generation algorithm Transformer-IPv6 for language modeling in the vector space. The new generation approach we used can render the language model obtaining creative sequences.
\item[3)] Experiments show that our approach outperformed conventional language models and state-of-the-art target generation algorithms on multiple metrics.
\end{itemize}

\textbf{Roadmap. } Sec. 2 summarizes the prior researches related to our work. Sec. 3 introduces the background of the IPv6 target generation. Sec. 4 and Sec. 5 highlights the overall design of IPv62Vec and Transformer-IPv6 components in 6VecLM. Sec. 6 shows the evaluation results and Sec. 7 concludes the paper.

\section{Related Work}
Prior work on IPv6 target generation falls into two broad categories:  (1) analyzing known addresses similarity to understand allocation patterns and (2) designing algorithms that generate candidate targets to scan. In addition, we will introduce (3) the related applications exploring semantic relationships.

\subsection{Address Similarity Learning}
To measure behavioral similarity among network hosts, Coull et al. \cite{coull2011measuring} proposed semantically meaningful metrics for common data types found within network data and compare its performance to a metric that ignores such information to underscore the utility. Ring et al. \cite{ring2017ip2vec} designed IP2Vec to learn the similarity of IP addresses. They used the meta-information about traffic as the context of the address to train the Word2Vec \cite{mikolov2013efficient} model. experiments demonstrate the effectiveness of clustering IP Addresses within a botnet data set. Our work is also based on Word2Vec to implement address similarity learning. However, active host discovery is the problem where meta-information is often lacking. We only rely on the active address set to discover new active hosts in this paper.

In the prior work of IPv6 active address set analysis, Planka et al. \cite{plonka2015temporal} first explored the potential patterns of IPv6 active addresses in time and space. They used Multi-Resolution Aggregate plots to quantify the correlation of each portion of an address to grouping addresses together into dense address space regions. Gasser et al. \cite{gasser2018clusters} employed entropy clustering to classify the hitlist into different addressing schemes. These efforts indicate that researchers have found a certain pattern hidden in the active IPv6 address sets, which provides a basis for the feasibility of IPv6 target generation.

\subsection{Target Generation Algorithm}
Ullrich et al. \cite{ullrich2015reconnaissance} used a recursive algorithm for the first attempt to address generation. They iteratively searched for the largest match between each bit of the address and the current address range until the undetermined bits were left, which is used to generate a range of addresses to be scanned. Murdock et al. \cite{murdock2017target} introduced 6Gen, which generates the densest address range cluster by combining the closest Hamming distance addresses in each iteration. Foremski et al. \cite{foremski2016entropy} used Entropy/IP for efficient address generation. They used a Bayesian network to model the statistical dependence between the values of different defined segments. This learned statistical model can then generate target addresses for scanning. Different from the previous approaches, our work tries to focus on the semantics of IPv6 for the first time to achieve the target generation algorithm through neural networks.

\subsection{Word Embedding and Language Modeling}
In order to explore the semantic relationship between words, Mikolov et al. \cite{mikolov2013efficient} proposed Word2Vec to learn high-quality distributed vector representations and prove the availability in measuring syntactic and semantic word similarities. With the development of word embeddings, Bengio et al. \cite{bengio2003neural} first employed neural networks to learn the joint probability function of sequences for substituting statistical language modeling, which subsequently led to deep learning gaining many successful experiences on language models \cite{mikolov2010recurrent,grave2016improving,dauphin2017language}. 

Recently, Vaswani et al. \cite{vaswani2017attention} proposed a completely self-attention-based network architecture Transformer. The model achieves state-of-the-art performance on the WMT 2014 English-to-German and English-to-French translation task. More work \cite{radford2018improving,devlin2018bert,al2019character,dai2019transformer} relies on the advantages of this model to achieve breakthroughs in applications. Our work is also based on the Transformer network. We modified the model to implement semantic discovery in the vector space.

\begin{figure}[t]
\includegraphics[width=\textwidth]{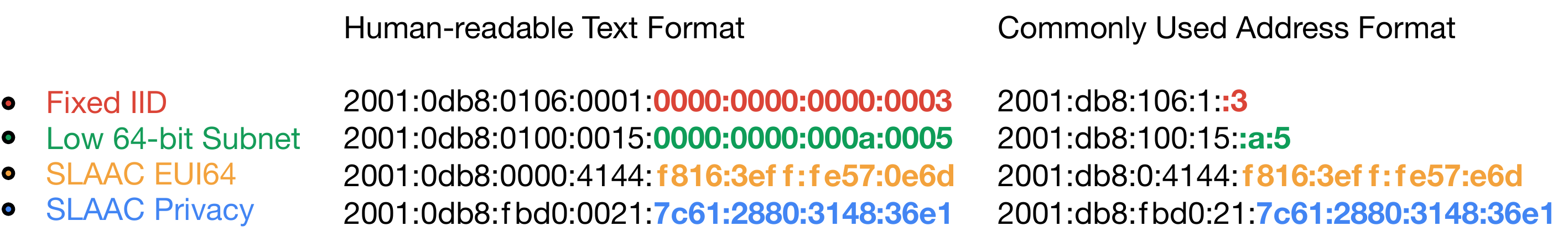}
\caption{Sample IPv6 addresses in presentation format with the low 64 bits shown bold.
} \label{fig1}
\end{figure}

\section{Preliminary}
While prior work of IPv6 analysis has obtained preliminary insights, active host discovery in the large and missing semantic IPv6 address space is still a huge challenge. In this section, we provide basic information on IPv6 and highlight our consideration of IPv6 target generation.

\subsection{IPv6 Addressing Background}
To explain IPv6 knowledge and the domain-specific terms we used in this paper, we provide a brief background on IPv6 addressing. We refer the reader to RFC 2460 \cite{deering2016internet} for a detailed description of the protocol.

An IPv6 address consists of a global network identifier, subnet prefix, and an interface identifier \cite{carpenter2014significance}. It is composed of 128-bit binary digits, which are usually represented in human-readable text format, using 8 groups of 4 hexadecimal digits and separating them by colons, as shown in Figure 1. Each of the hexadecimal digits is called a nybble. IPv6 addresses usually use ”::” to replace groups of consecutive zero values and omit the first zero value in each group.

However, IPv6 addresses are not simply composed of meaningless digits. There are many IPv6 addressing schemes and network operators are reminded to treat interface identifiers as semantically opaque \cite{carpenter2014significance}. Administrators have the option to use various standards to customize the address types. In addition, some IPv6 addresses have SLAAC \cite{thomson1998ipv6} address format that the 64-bit IID usually em- beds the MAC address according to the EUI-64 standard \cite{thomson1998ipv6} or uses completely pseudo-random \cite{narten2001privacy}. Consider the sample addresses in Figure 1. In increasing order of complexity, these addresses appear to be: (1) an address with fixed IID value (::3). (2) an address with a structured value in the low 64 bits (perhaps a subnet distinguished by :a). (3) a SLAAC address with EUI-64 Ethernet-MAC-based IID (ff:fe flag). (4) a SLAAC privacy address with a pseudorandom IID.

\subsection{Target Generation Consideration}

\subsubsection{IPv6 Address Space}
Active hosts are scattered in the sparse IPv6 space because of excessive address reserves. Limited by computational power, a brute-force approach to probe the entire network space of IPv6 is almost impossible. The distribution of active addresses is also difficult to extract. Therefore, we consider constructing a vector space with good interpretability, where the distance between vectors can be defined as the relationship between addresses. However, IPv6's large address range means that even hexadecimal IPv6 addresses have 32 nybbles. It is difficult to build a high-quality representation of an address vector in a high-dimensional space. Our work will focus on address space representation through model learning and utilize dimensionality reduction techniques to obtain active address clustering areas.

\subsubsection{IPv6 Semantic}
It may be difficult to effectively train the learning model when analyzing the structure of the address set due to the opaque semantics of IPv6 addresses and the existence of multiple addressing schemes. In order to design an effective target generation algorithm, we believe that reasonably mining the semantic information of address composition is particularly critical. We define the IPv6 semantics by building sequences of address words. The context of the address word sequences can be learned through a model to generate the address vectors with semantic discrimination, which contributes to learning address word sequence relationships to speculate address composition by language modeling.

\begin{figure}[t]
\begin{center}
\includegraphics[width=0.9\textwidth]{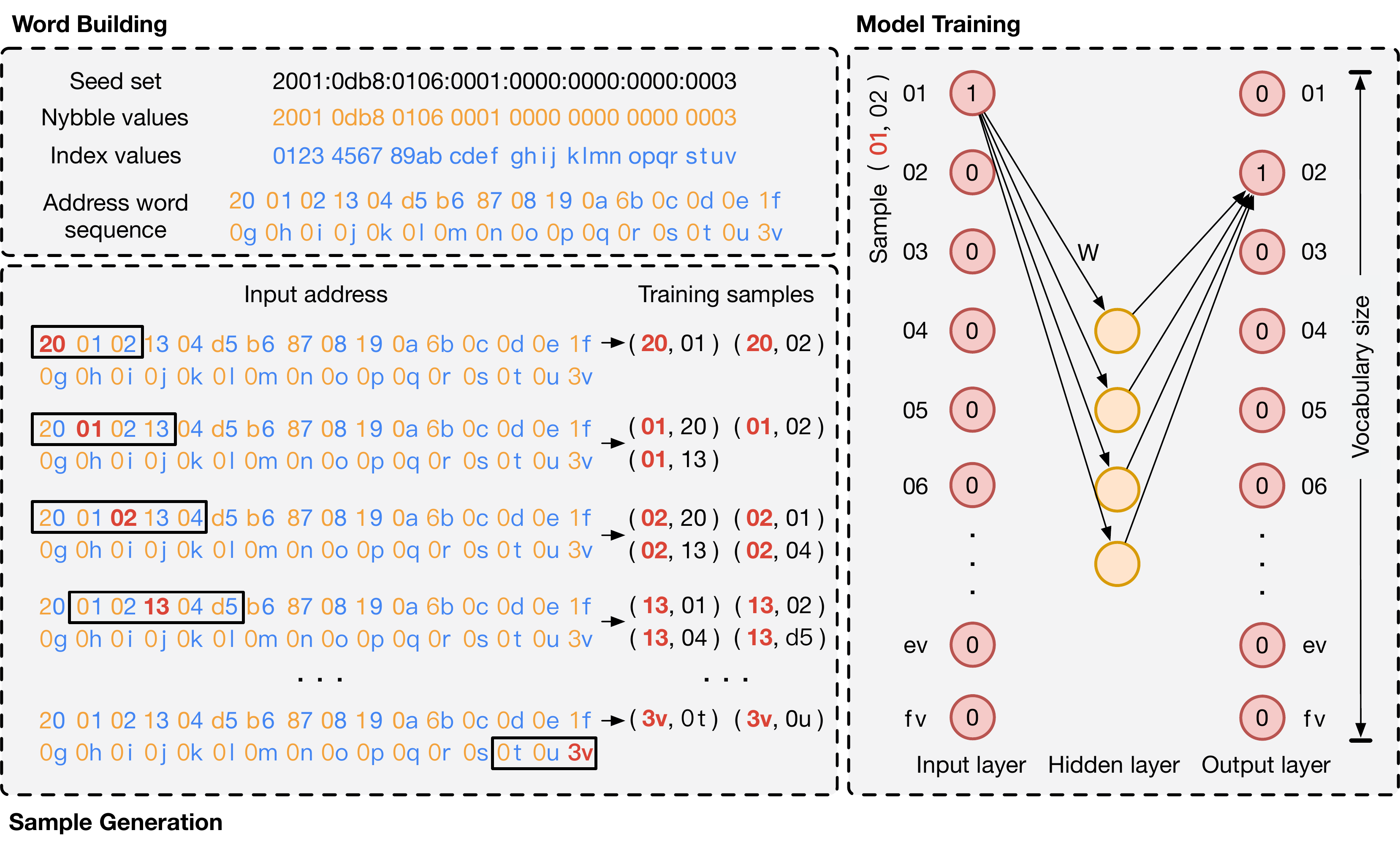}
\caption{The overall architecture of IPv62Vec. An address word sequence is composed of the nybble and index values of an address from the seed set. The training samples are generated from the corresponding combinations of input words (highlighted in red color) and context words. The neural network is trained with samples and outputs the vector representation W of the input word.
} \label{fig2}
\end{center}
\end{figure}

\section{IPv62Vec}
In this section, we will introduce our first component in 6VecLM, IPv6 vector space mapping technology IPv62Vec. We outline the underlying ideas of the work including word building, sample generation, and model training.

\subsection{Word Building}
Constructing effective semantic information requires to define a new semantic representation of the address. In Figure 2, we re-represent each nybble of the hex address to create address words. We define the value $V_i$ of the $i$-th nybble in an address, where $V\in\{0,1,...,f\}$. The index $i$ is defined as $S_i$, where $S\in\{0,1,...,v\}$. The $i$-th address word in a new representation is composed of the nybble value and index value as $V_iS_i$ (e.g. the 11th nybble value 2 is represented as the address word 2a). All address words built from the seed set is defined as vocabulary. Our purpose is to distinguish the nybble values at different indexes. We consider that the same nybble values usually have different degrees of semantic importance according to their position in an address. Differentiating work contributes to discovering the semantic information of key positions (e.g. The 23rd-26th nybbles of the SLAAC EUI-64 address—fffe).

\subsection{Sample Generation}
After determining the address words, we follow the word selection process of Mikolov et al. \cite{mikolov2013efficient} to select input words and context to generate training samples. As shown in Figure 2, we perform a word selection operation on each address word sequence in the seed set. When a certain word of the sequence is selected as an input word, words from the surrounding window of the input word are chosen as context words for building training samples. The window size is 5.

\subsection{Model Training}
Since neural networks cannot be fed with words, each word is represented as a One-Hot vector and the length of this vector is equal to the size of the vocabulary. The number of input and output neurons of the neural network is equal to the size of the vocabulary. Further, the output layer uses a softmax classifier and indicates the probabilities that a particular word appears in a specific context. The neural network is fed with the input word and tries to predict the probability of the context word. The output layer of the neural network indicates how likely each word of the vocabulary may be found in the context of the input word. After training, the final hidden layer result is the vector representation of the input word. In this paper, we used 100 neurons in our hidden layer.

\begin{figure}[t]
\begin{center}
\includegraphics[width=0.9\textwidth]{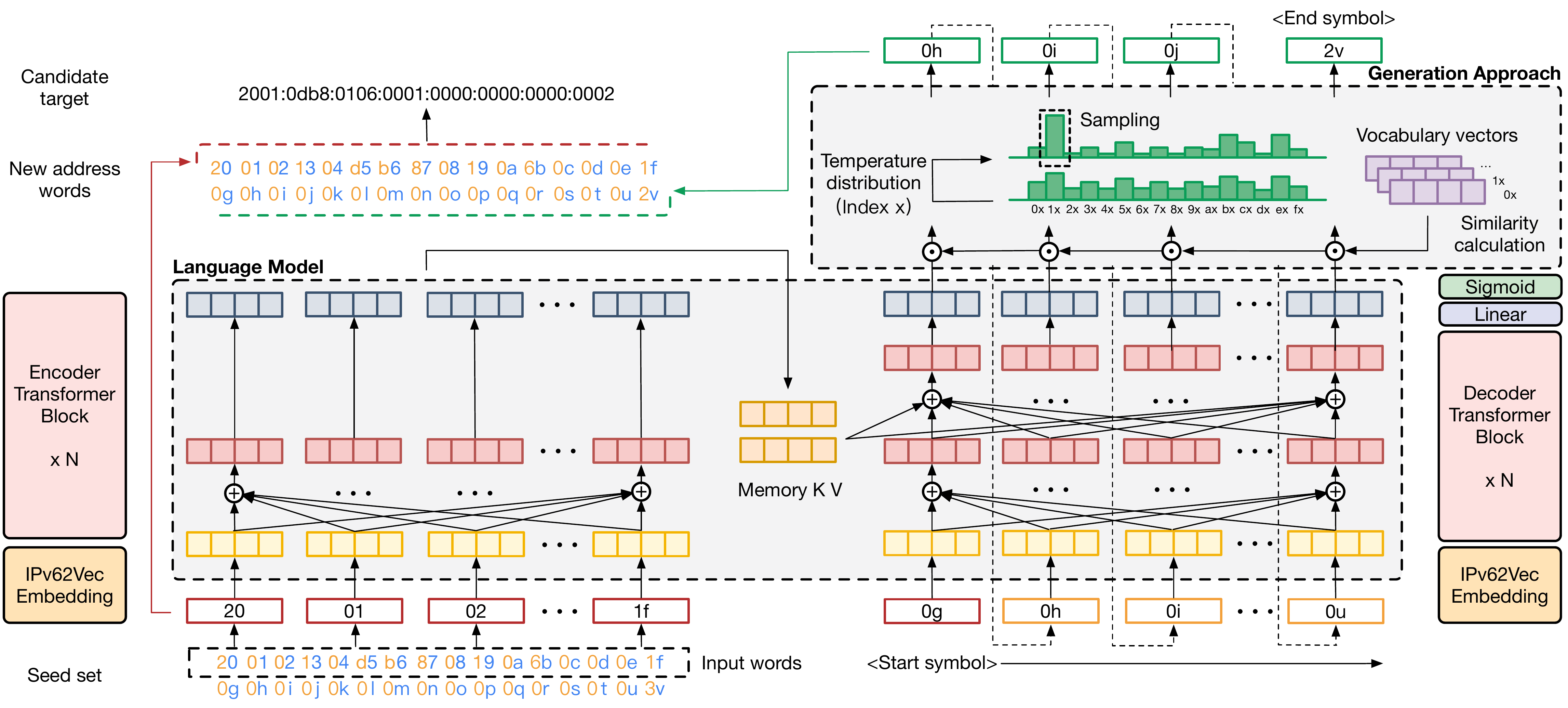}
\caption{The overall architecture of Transformer-IPv6. The language model based on the Transformer network can generate address word vector by learning past words. Then the new generation approach will generate sampling probability distribution based on cosine similarity and adjust the distribution by softmax temperature. The output words are recombined with the input to generate a new candidate target.
} \label{fig3}
\end{center}
\end{figure}

\section{Transformer-IPv6}
After obtaining the IPv6 address vector, the next step is how to apply this semantic vector space to complete network-related tasks. In this section, we will employ the address vector to realize the IPv6 target generation through building the second component in 6VecLM, our language model Transformer-IPv6.

\subsection{Language Modeling}
Language models assign a probability distribution over sequences $t_{0:L}$ by factoring out the joint probability as follows, where $L$ is the sequence length:
\begin{equation} 
P(t_{0:L})=P(t_0)\prod_{i=1}^LP(t_i|t_{0:i-1})
\end{equation}
To model the conditional probability $P(t_i|t_{0:i-1})$, we train a Transformer network to process the address word sequence $t_{0:i-1}$. Transformer is a self-attention-based deep network. We consider focusing on the semantic importance of address word in the sequence for modeling the probability of the next word given the preceding word, until obtaining an entire address word sequence.

The architecture of Transformer-IPv6 is shown in Figure 3. The input words are converted from addresses in the seed set according to the word building method in Sec. 4. The vector representation, which is determined by the pre-trained IPv62Vec, of the first 16 words of the sequence is inputted in the model to predict the last 16 words. The model then stacks $n$ layers of Transformer encoder block to encode latent vector as a memory. Following Vaswani et al. \cite{vaswani2017attention}, the Transformer encoder block contains a multi-head self-attention sub-layer followed by a feed-forward network of two fully connected sub-layers. A residual connection and layer normalization are followed each of the two sub-layers. The attention mechanism can help addresses consider critical parts of the sequence when predicting words, while the multi-head attention mechanism observes more address word combinations by training multiple attentions:
\begin{equation} 
\begin{aligned}
{\rm Attention}(Q,K,V) &= {\rm softmax}(\frac{QK^T}{\sqrt{d_k}})V\\
\end{aligned}
\end{equation}
\begin{equation} 
\begin{aligned}
{\rm MultiHead}(Q,K,V) &= {\rm Concat}({\rm head}_1,...,{\rm head}_h)W^O\\
{\rm where}\ {\rm head}_i &= {\rm Attention}(QW^Q_i,KW^K_i,VW^V_i)
\end{aligned}
\end{equation}
Where $Q, K, V$ is the output of the upper layer. $ d_q $, $ d_k $, $ d_v $, $ d_ {\rm model} $ are the dimensions of the matrix $ W ^ Q_i $, $ W ^ K_i $, $ W ^ V_i $, and the model input. The linear projections are parameter matrices $W^Q_i\in\mathbb{R}^{d_{\rm model}\times d_q}$, $W^K_i\in\mathbb{R}^{d_{\rm model}\times d_k}$, $W^V_i\in\mathbb{R}^{d_{\rm model}\times d_v}$, $W^O_i\in\mathbb{R}^{d_{\rm model}\times hd_v}$. 

The last 16 words use the mask method \cite{vaswani2017attention} to select the current input of the Transformer decoder and ensure that the model’s predictions are only conditioned on past words. Transformer decoder block inserts a second multi-head self-attention sub-layer, which performs attention weights computation while keeping encoder memory as attention input $K$ and $V$.

The model finally predicts the next address word vector through a linear layer and a sigmoid activation function until completing an entire address generation process. Our model uses Transformer block layers $ n = 6 $, attention head numbers $ h = 10 $, parameter matrix dimension $ d_q = d_k = d_v = 10 $, and model input dimension $ d_ {\rm model} = 100 $.

\subsection{Generation Approach}
In order to complete the target generation task in the vector space, We expect the generated word vector $ y_ {\rm pred} $ to have a high semantic similarity to the target word vector $ y_ {\rm true} $. Therefore, our model uses the cosine distance as the loss function $L$:
\begin{equation}
\begin{aligned}
{\rm cos}(\theta) &= \frac{\sum^n_{i=1}y^{(i)}_{\rm true}\cdot y^{(i)}_{\rm pred}}{\sqrt{\sum^n_{i=1}(y^{(i)}_{\rm true})^2}\cdot \sqrt{\sum^n_{i=1}(y^{(i)}_{\rm pred})^2}}\\
L &= 1 - {\rm cos}(\theta)
\end{aligned}
\end{equation}
Unlike conventional language models that directly model word probabilities, our approach predicts word vectors to preserve the semantic information of the vector space. Since our training samples are address vectors with semantic relationships obtained by IPv62Vec, minimizing the cosine distance can obtain prediction targets with similar context structure to the seed address. This approach aims to choose the closest address word in the vector space, which contributes to discovering the active addresses cluster area.

After generating the address word vector in each epoch, we calculate the cosine similarity between the predicted word vector and each word vector containing the current index in the vocabulary, which is used as a basis for sampling the predicted words. We employ the softmax function to convert the cosine similarity ${\rm cos}(\theta)$ to the word sampling probability $P(i)$:
\begin{equation}
P(i) = \frac{e^{{\rm cos}(\theta)_i}}{\sum^C_{j=1}e^{{\rm cos}(\theta)_j}}\ ,\ i=1,...,C
\end{equation}
Where $C$ is the number of words with the current index in the vocabulary.

To build an effective generative model, we consider two word sampling strategies: greedy sampling and random sampling. The greedy sampling selects the word with the highest sampling probability in each epoch, while the generated address is always similar to the training set address and raises a high repetition rate. Random sampling ignores the sampling probability and always randomly selected words, while the generated address has high randomness and excessively loses semantics, thus leading to a low activity rate. To seek a balance between maintaining semantics and creativity, we use softmax temperature \cite{muller2019does} to readjust the probability distribution:
\begin{equation}
Pr(i) = \frac{e^{{\rm log}P(i)^{1/t}}}{\sum^C_{j=1}e^{{\rm log}P(j)^{1/t}}}\ ,\ i=1,...,C
\end{equation}
Where temperature $t$ is a hyperparameter. A high temperature $t$ leads close sampling probability of each word, thus the sampled address is more random. While a low temperature $t$ enhances the difference of original sampling probability, which results in a strong ordering of the generated address.

\section{Evaluation}
In this section, we evaluate the performance of our approach. We will introduce the data set and evaluation method used in the experiment and show the effectiveness of our approach on the active address set.

\begin{table}

\caption{The detail of the two active address datasets we used in the paper.}\label{tab2}
\begin{center} 
\begin{tabular}{|c|c|c|c|}
\hline
Dataset & Seeds & Period & Collection Method\\
\hline
\hline
IPv6 Hitlist & 100,000 & January 9, 2020  & Public\\
CERN IPv6 2018 &  90,010 & March 2018 - July 2018 & Passive measurement\\
\hline
\end{tabular}
\end{center} 
\end{table}

\subsection{Dataset}
Our experimental datasets are mainly from two parts, a daily updated public dataset IPv6 Hitlist and a measurement dataset CERN IPv6 2018. Table 1 summarizes the datasets used in this paper. The public dataset IPv6 Hitlist is from the data scanning the IPv6 public list for daily active addresses, which is provided by Gasser et al \cite{gasser2018clusters}. In addition, we passively collected address sets under the China Education and Research Network from March to July 2018. We continued to scan and track the IPs that are still keeping active as our measurement dataset CERN IPv6 2018.

\subsection{Evaluation Method}

\subsubsection{Scanning Method}
To evaluate the activity of the generated address, we use the Zmapv6 tool  \cite{durumeric2013zmap}  to perform ICMPv6, TCP/80, TCP/443, UDP/53, UDP/443 scans on the generated address. When the query sent by any scanning method gets a response, we will determine the address as active. Noting the difference in activity between hosts at different times, we maintain continuous scanning of the host for 3 days to ensure the accuracy of our method.

\subsubsection{Evaluation Metric}
Since IPv6 target generation is different from text generation tasks, we need to define a new evaluation metric for the address generative model. In the case of a given seed set, $N_{candidate}$ represents the number of the generated candidate set, $N_{hit}$ represents the number of generated active addresses, $N_{gen}$ represents the generated address that is active and not in the seed set. Then the active hit rate $r_{hit}$ and active generation rate $r_{gen}$ of the model can be computed as 
\begin{equation}
r_{hit} = \frac{N_{hit}}{N_{candidate}} \times 100\% \qquad \qquad r_{gen} = \frac{N_{gen}}{N_{candidate}}  \times 100\%
\end{equation}
$ r_ {hit} $ can represent the model's learning ability to learn from the seed set. $ r_ {gen} $ highlights the model's generation ability to generate new active addresses. 

\subsection{IPv6 Vector Space}
To illustrate the effectiveness of our approach IPv62Vec, we use the active address set IPv6 Hitlist and construct training samples as described in Sec. 4. After training the model, we extract the hidden layer parameters of the model to build the mapping relationship between the address word and the word vector.

\begin{figure}[t]
\begin{center}
\includegraphics[width=0.9\textwidth]{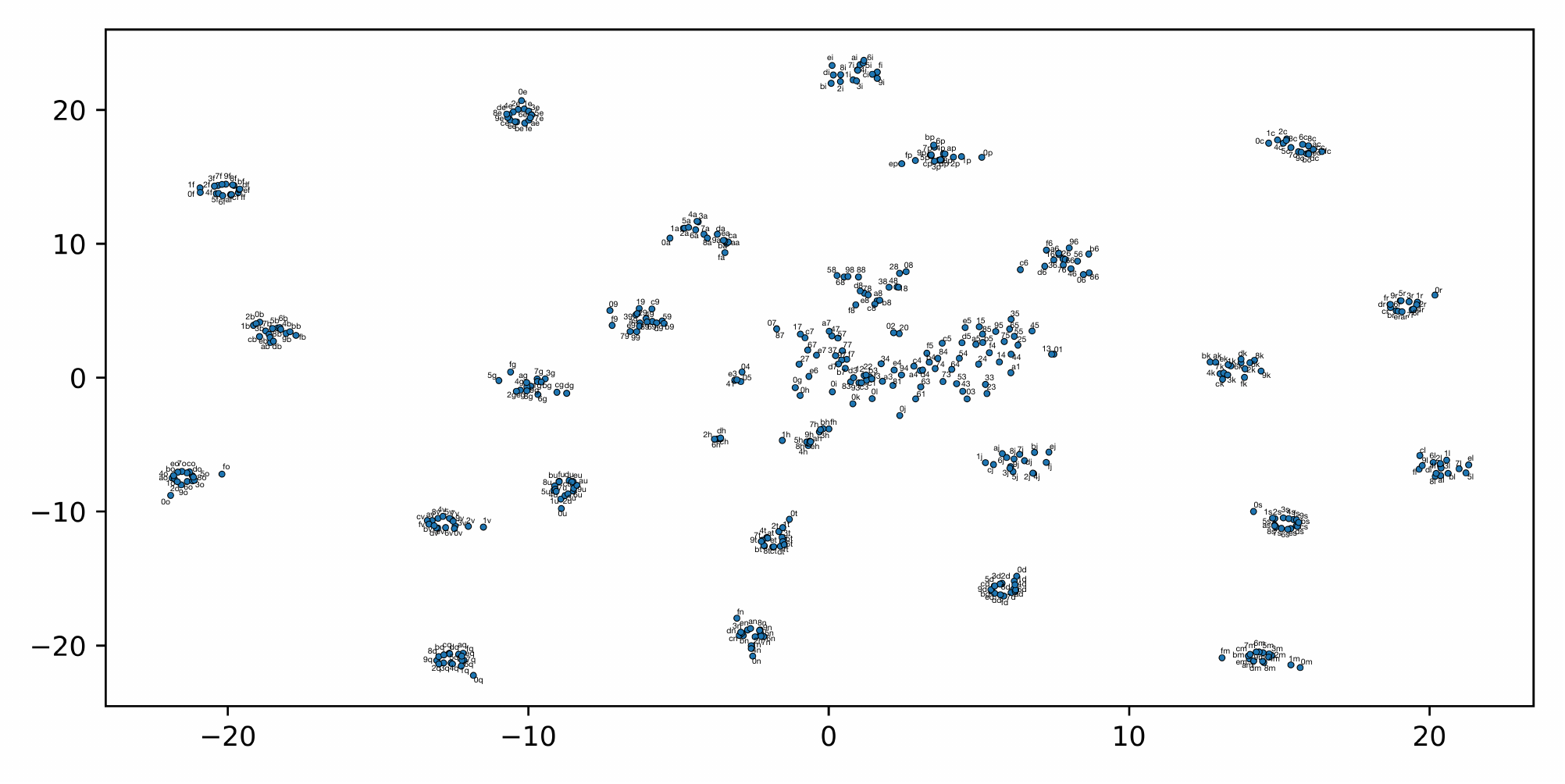}
\caption{The address word distribution by using t-SNE visualization. The words with a similar context in an address are clustered in the vector space.
} \label{fig4}
\end{center}
\end{figure}

\subsubsection{Word Vector Space}
Since address word vectors have high dimensions, we use t-SNE technology \cite{maaten2008visualizing} to reduce the dimensionality of word vectors to facilitate display. Figure 4 shows the semantic relationship between address words in word vector space. Address words with similar contexts perform a tight cluster in vector space. We found that most address words are clustered according to their index attributes, which indicates that different nybble values with the same index possess similar contexts. While the long distance between address words with the same nybble value indicates that they keep different contexts with different indexes, which confirms our intention on word building. Nybble value 0 is an exception to this proposition because address words with nybble value 0 at some index perform a certain degree of clusters, such as index value 4-5, g-l. We surmise that consecutive zeros in the address are the reason for this situation. In addition, the address words with index values 0-7 are close, which indicates that the network prefixes of addresses often have similar structures.

\subsubsection{Address Vector Space}
The address vectors are determined by combining the address word vectors contained in each address in the address set. In addition, we use a One-Hot vector to represent the characters 0-f and construct the address vector according to the active addresses composition as a baseline for comparison. We use t-SNE technology to reduce the dimension of the address vectors and employ DBSCAN \cite{ester1996density} to complete the unsupervised clustering for display. Figure 5 shows the address distribution in the vector space. IPv62Vec successfully divided addresses into several classes. Addresses under the same class perform a high similarity. The One-Hot address vector cannot mine the addresses similarity due to the lack of semantic information.

IPv62Vec only relies on the address sequences to perform effective address similarity learning. The network prefix, subnet identifier, and interface identifier in an address cannot be determined due to the opaque sequence. IPv62Vec extracts the potential network features, thus performing an effective address cluster. By extracting features on IPv6 addresses, we consider that the approach will also be feasible for other network tasks, such as encrypted traffic classification.

\begin{figure}[t]
\centering
\subfigure[One-Hot]{       
\label{1}
\begin{minipage}[t]{0.47\linewidth}
\centering
\includegraphics[width=5.8cm]{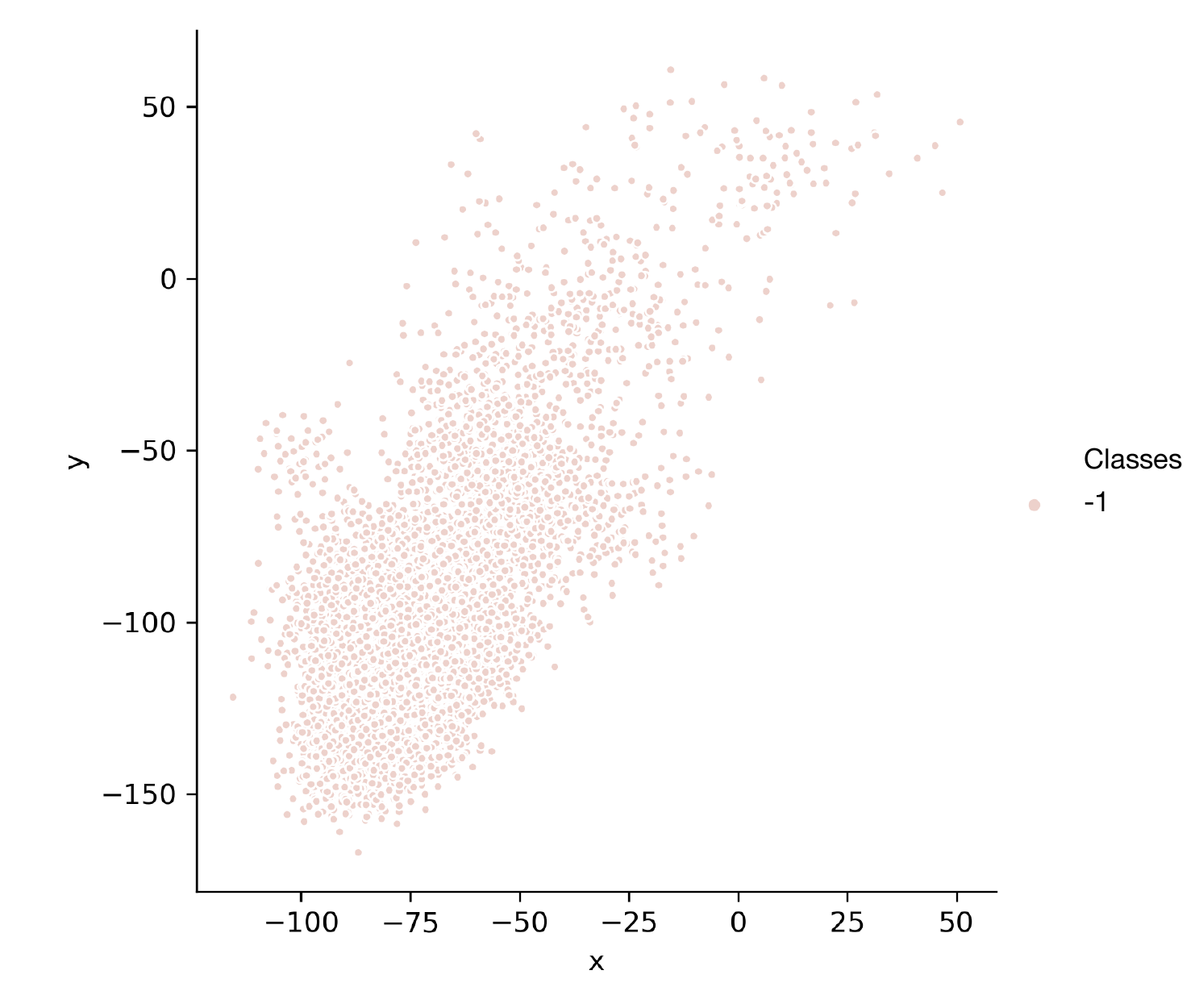}
\end{minipage}
}
\subfigure[IPv62Vec]{ 
\label{2}
\begin{minipage}[t]{0.47\linewidth}
\centering

\includegraphics[width=5.8cm]{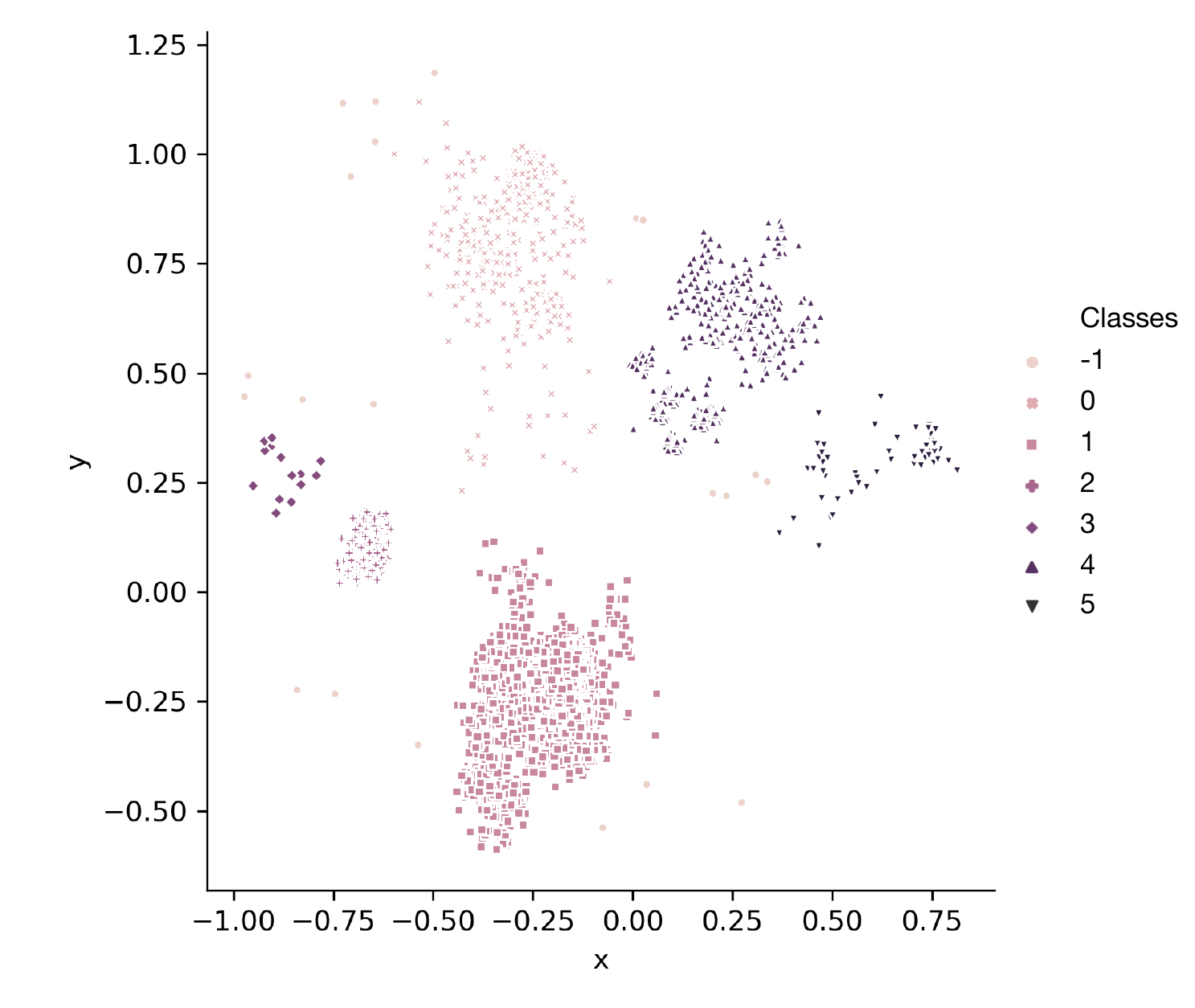}
\end{minipage}
}
\centering
\caption{Comparison of address vector distribution between Ont-Hot Encoding and IPv62Vec. IPv62Vec classified the address set to 6 categories by learning the address similarity on the data set IPv6 Hitlist.}
\label{fig5}
\end{figure}

\begin{figure}[t]

\includegraphics[width=\textwidth]{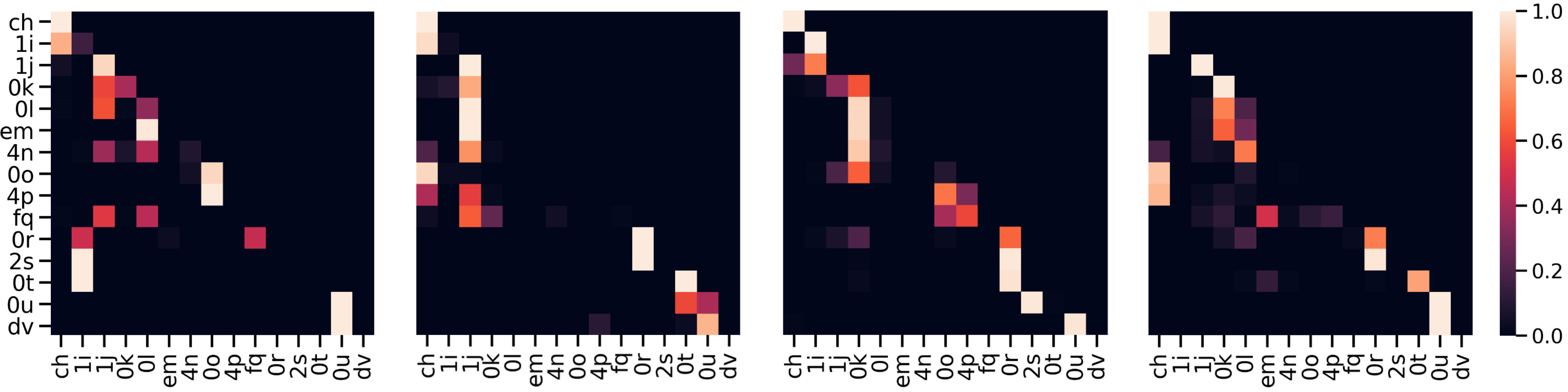}
\caption{An example of 4 attention heads on address word sequence in the decoder self-attention in layer 5 of 6. Transformer-IPv6 can conclude a comprehensive target by considering multiple attention results.
} \label{fig6}
\end{figure}

\subsection{Address Attention}
After performing the address vector space mapping, we use Transformer-IPv6 to learn the address vector. To illustrate the effectiveness of Transformer-IPv6, we performed attention visualization work to show the model performance. Figure 6 shows the model's focus on IPv6 addresses. In IPv6 address words, each word has a clear object of attention in the entire address, which enables the associated information in the address to be effectively mined. For example, 1i, 1j, 0k, and 0l may respectively have strong correlations in Figure 6. In addition, the multi-head attention mechanism guarantees the diversity of address attention, which renders the final address word output of the model to integrate multiple possibilities.

\begin{figure}[t]
\begin{center}

\includegraphics[width=0.9\textwidth]{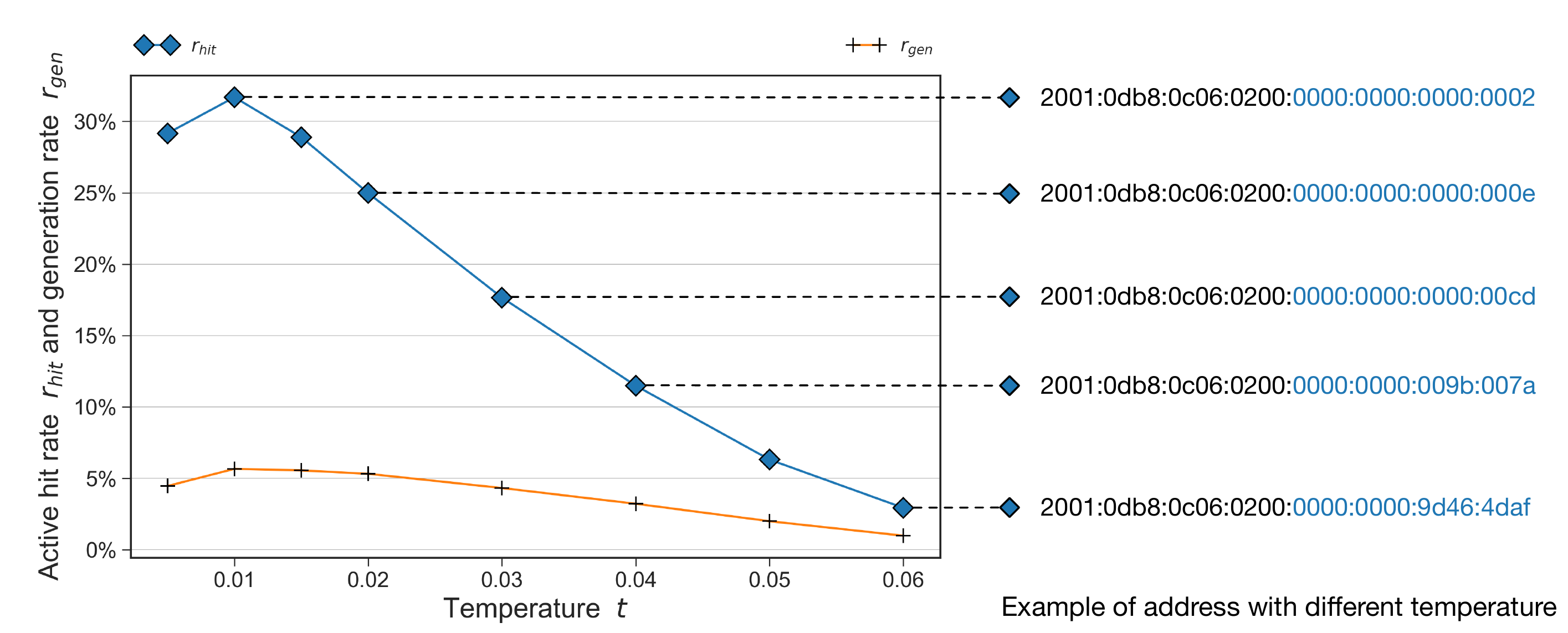}

\caption{The predicted address results with different softmax temperature $t$. Increasing temperature $t$ will generate targets with more random address words. 
} \label{fig7}
\end{center}
\end{figure}

\subsection{Temperature}
In our model, softmax temperature is a key parameter that can control the quality of the addresses generated by the model. When selecting a high temperature $ t $, the model tends to sample randomly and the generated address contains more creative sequences. The model is required to sample greedily and the generated address is more close to the seed set when keeping a low temperature $ t $. Figure 7 shows the generation results corresponding to different temperature $ t $. The increase of temperature promotes the generation of address words more diversified. In order to find the equilibrium point, we measured the generation performance corresponding to different $t$ values. The model keeps the highest active hit rate $ r_ {hit} $ and active generation rate $ r_ {gen} $ when $t = 0.01$. We recommend $t$ value between 0-0.05 to ensure the model performance.

\subsection{Evaluation Results}

\subsubsection{Baselines} 
The baselines in our experiments for comparison mainly contain: (1) conventional language model. RNN \cite{mikolov2010recurrent}, LSTM \cite{grave2016improving} and GCNN \cite{dauphin2017language} are the prior paradigms that have shown significant gains in language modeling. In addition, we added IPv62Vec and our generation approach to the conventional language model for adapting the model to target generation tasks. (2) target generation algorithm. Entropy/IP \cite{foremski2016entropy} and 6Gen \cite{murdock2017target} are the state-of-the-art address generation tools that can also efficiently generate active IPv6 targets. We employed the open-source code of Entropy/IP and implemented 6Gen according to the algorithm described by the authors to build  baselines.

\subsubsection{Experimental Results}
Table 2 demonstrates the performance of all the compared models based on the public data set IPv6 Hitlist. The results show 6VecLM outperforms all the baselines, which confirms the advantage of the IPv62Vec and Transformer-IPv6 mechanism. Entropy/IP and 6Gen performs poorly compared to our approach due to lacking IPv6 semantics and adaptability to data sets. By adding generation approach and IPv62Vec mechanism, the conventional language models can reach a not bad performance. While our model is more competent to the target generation task due to the multiple address attention mechanism in Transformer-IPv6.

\begin{table}[t]
\caption{The experimental results by comparing with conventional language models and target generation algorithms Entropy/IP and 6Gen. Results show that 6VecLM reached the best performance in our experiments.}\label{tab5}
\begin{center} 
\begin{tabular}{|c|c|c|c|c|c|c|}
\hline
Category & Model & $N_{candidate}$ & $N_{hit}$ &  $N_{gen}$ & $r_{hit}$ & $r_{gen}$\\
\hline
\hline
Conventional&RNN \cite{mikolov2010recurrent}&34,604&995&851&2.88\%&2.46\%\\
language&LSTM \cite{grave2016improving}&34,636&727&564&2.10\%&1.63\%\\
model&GCNN \cite{dauphin2017language}&34,817&787&649&2.26\%&1.86\%\\
\hline
Target generation&Entropy/IP \cite{foremski2016entropy} & 69,167 & 8,321 & 2,540 & 12.03\% & 3.67\%\\
algorithm&6Gen \cite{murdock2017target} & 67,712 & 4,612 & 1,638 & 6.81\% & 2.42\%\\
\hline
Adding&RNN \cite{mikolov2010recurrent} & 44,242 & 12,133 & 2,409 & 27.42\% & 5.44\%\\
IPv62Vec and&LSTM \cite{grave2016improving} & 61,950 & 10,640 & 2,019 &  17.18\% & 3.26\%\\
generation approach&GCNN \cite{dauphin2017language} & 52,046 & 11,360 & 2,146 & 21.83\% & 4.12\%\\
\hline
Our approach& 6VecLM & 46,461 & {\bfseries 15,406} & {\bfseries 2,883} & {\bfseries 33.16\%} & {\bfseries 6.21\%}\\
\hline
\end{tabular}
\end{center} 

\end{table}

\begin{figure}[t]

\centering
\subfigure[IPv6 Hitlist]{       
\label{1}
\begin{minipage}[t]{0.47\linewidth}
\centering
\includegraphics[width=5.9cm]{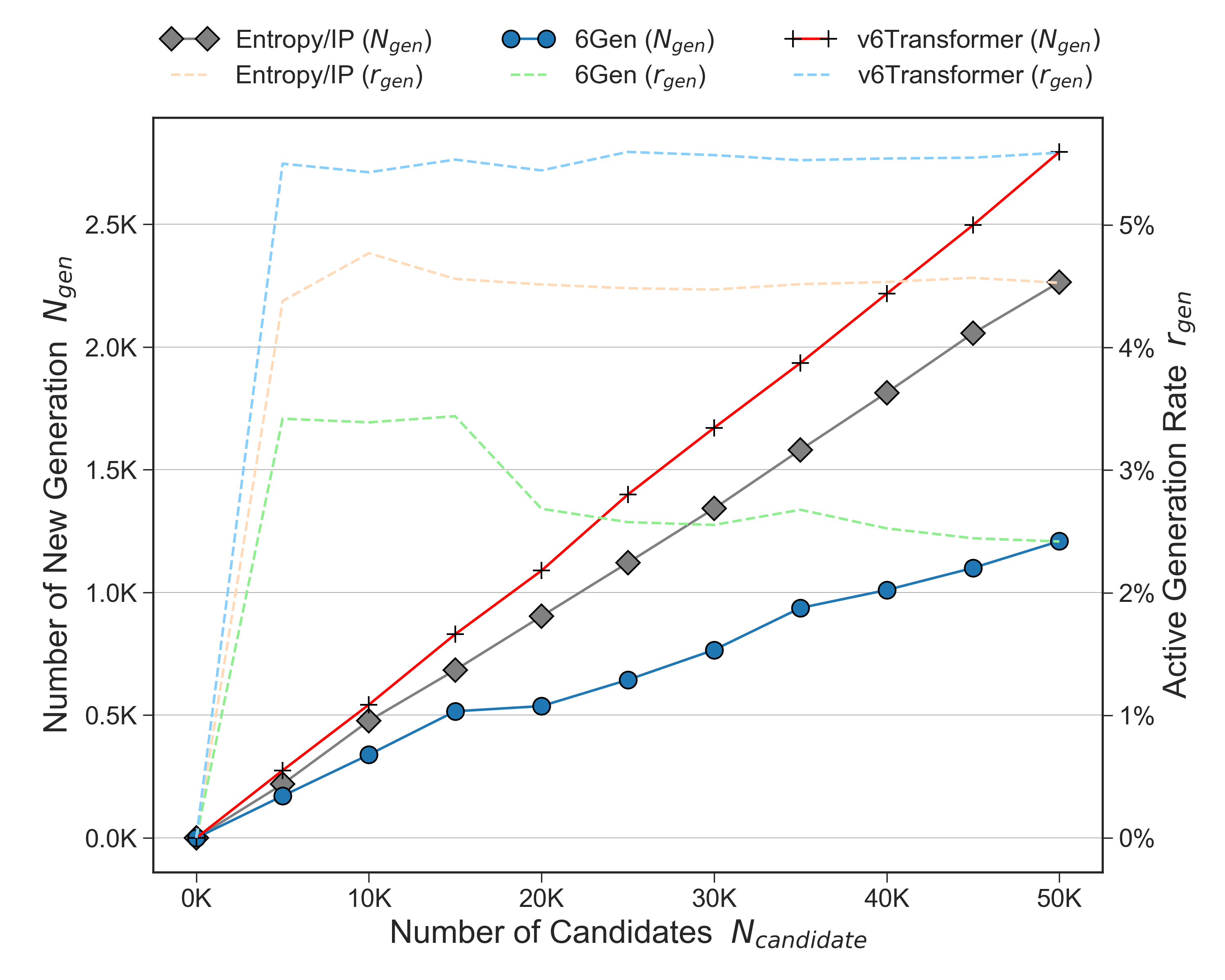}
\end{minipage}
}
\subfigure[CERN IPv6 2018]{ 
\label{2}
\begin{minipage}[t]{0.47\linewidth}
\centering

\includegraphics[width=5.9cm]{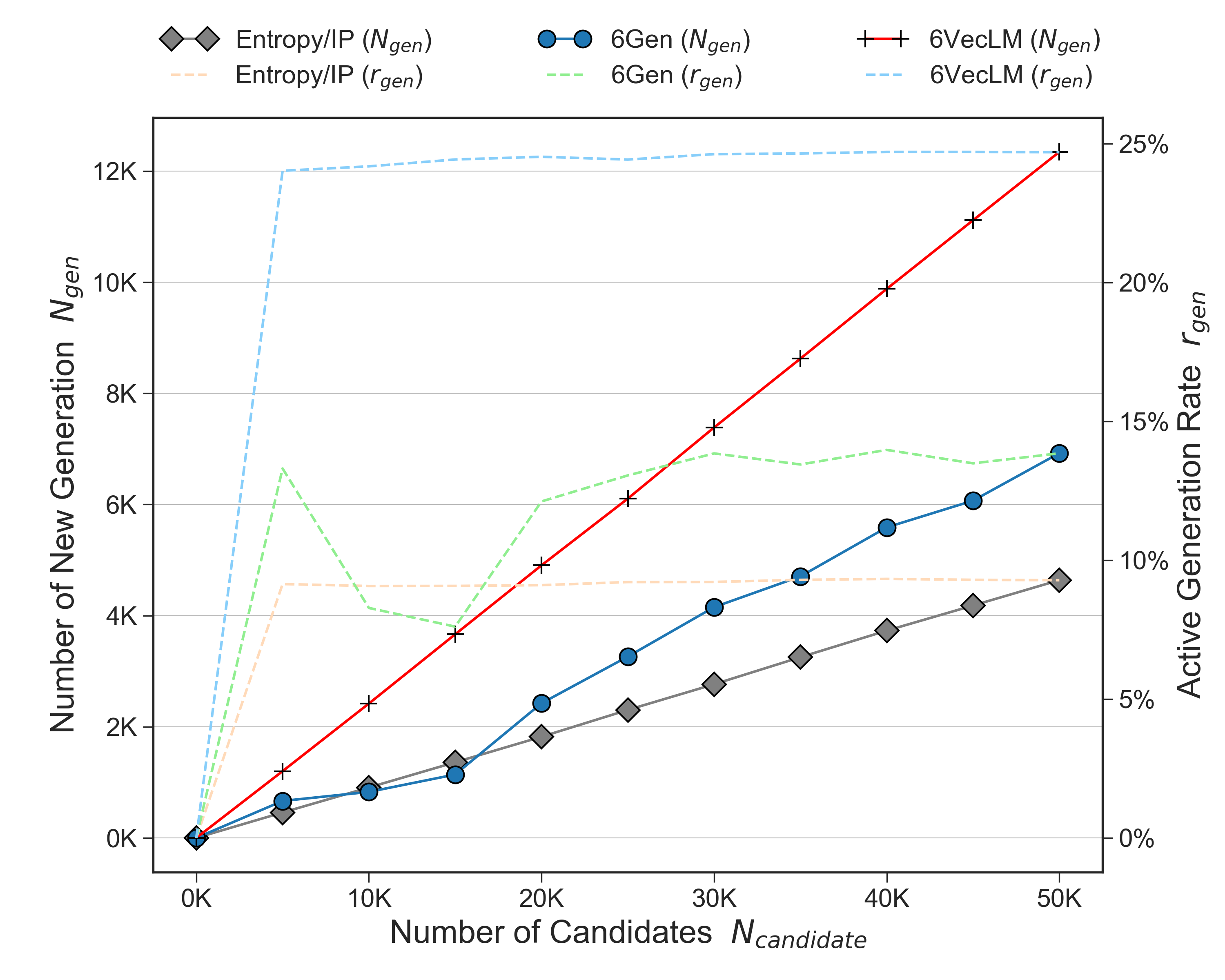}
\end{minipage}
}
\centering
\caption{The experimental results by comparing with Entropy/IP and 6Gen on the two datasets. $N_{gen}$ and $r_{gen}$ are evaluated under the different candidate set $N_{candidate}$.}
\label{fig8}
\end{figure}

\subsection{Generating Ability}
In order to evaluate the model's generating ability, in Figure 8, we tested the number of new active generation $ N_ {gen} $ and the active generation rate $ r_ {gen} $ under the different number of generated candidates on the two address sets. The performance of Entropy/IP and 6Gen is slightly different under the two data sets. Because of the algorithm's ability to adapt to different data sets, the data set CERN IPv6 2018 may have denser address areas than IPv6 Hitlist, which is more conducive to 6Gen algorithm. However, experiment results indicate that our approach reaches a better performance than the other two algorithms. 6VecLM can find 1.23-2.66 times and 1.78-2.31 times more hits than Entropy/IP and 6Gen. The model has a stable and good performance under different data sets. We expect that future target generation algorithms will be capable of generating more valid targets under a limited size of the candidate set to ensure predicting high-quality candidate sets.

\section{Conclusion}
In this work, we explored the basic challenge of generating promising IPv6 addresses to scan.  We presented 6VecLM, an approach to map addresses to a vector space and implement an IPv6 language model that can generate addresses. The address vector generated by IPv62Vec mechanism in 6VecLM effectively extracts the underlying semantic information of the address. Transformer-IPv6 mechanism can learn the word sequences in the vector space and select the address generation strategy relying on the cosine similarity and softmax temperature. The work is superior to conventional language models and state-of-the-art target generation algorithms Entropy/IP and 6Gen. 

\subsubsection{Acknowledgements}
This work is supported by The National Key Research and Development Program of China (No.2016QY05X1000) and The National Natural Science Foundation of China (No. U1636217) and The National Key Research and Development Program of China (No. 2020YFE0200500 and No. 2018YFB1800200).
%
%
%

\end{document}